\documentclass[%
 reprint,
 amsmath,amssymb,
 aps,
]{revtex4-2}
\usepackage{graphicx}
\usepackage{tabularx}
\usepackage{epsfig}
\usepackage{comment}
\usepackage{amsfonts,color}
\usepackage{dcolumn}
\usepackage{bm}

\begin{document}

\preprint{APS/123-QED}

\title{Impact of virtual REU experiences on students' psychosocial gains during the COVID-19 pandemic}

\author{Dina Zohrabi Alaee}
\email{dxzsps@rit.edu}
\author{Micah K. Campbell}%
 
 \author{Benjamin M. Zwickl}%
\affiliation{%
School of Physics and Astronomy, Rochester Institute of Technology, Rochester, New York 14623, USA 
}%
\date{\today}

\begin{abstract}
In the Summer of 2020, due to COVID-19, institutions either canceled or remotely hosted their Research Experience for Undergraduates (REU) programs. We carried out a 16-week longitudinal study examining the impact of these fully remote research experiences on mentees' psychosocial gains (e.g., identity). We studied the phenomenon of a remote research experience from the standpoint of the mentees ($N$=10) and their mentors ($N$=8), who were each interviewed seven and three times, respectively (94 total interviews). 
All mentees reported that this experience was highly beneficial through different factors and that they developed a sense of belonging, self-efficacy, and identity despite working remotely. Then, we synthesize these results with prior literature and develop a framework showing how different experiences and constructs affect the development of the physics and researcher identity. Gaining greater understanding regarding factors leading to the growth of psychosocial gains may help REU coordinators and REU mentors to re-design their undergraduate research program and provide the support that their mentees' needed to have a positive undergraduate research experience. 
\end{abstract}

\maketitle

\section{\label{sec:Intro}Introduction}

Undergraduate research is an important experience that affects student education and career development \cite{council_transforming_1999, council_improving_2002, wenzel_enhancing_2004, boyer_boyer_2003, kuh_high-impact_2008}. Across many studies, undergraduate research programs that include in-person interactions between the research mentor and mentee, access to physical equipment, and supportive lab environments have been documented as beneficial \cite{kuh_high-impact_2008, mabrouk_student_2000, pascarella_how_2005}. Substantial prior research has examined the influence of in-person undergraduate research experiences (UREs) on promoting outcomes, such as retention \cite{russell_pipeline_2007, cuthbert_it_2012, nagda_undergraduate_1998, shanahan_playing_2007, rodenbusch_early_2016, bauer_alumni_2003, hathaway_relationship_2002, lopatto_survey_2004, adedokun_understanding_2012, hunter_becoming_2007, seymour_establishing_2004}, and persistence \cite{chemers_role_2011, melanie_t_jones_importance_2010, schultz_patching_2011, pascarella_what_1996} in STEM career pathways. For instance, Bauer and Bennett \cite{bauer_alumni_2003} contend that undergraduate research may help mentees clarify future career goals and are more likely to attend graduate school. 

Beyond the benefits of undergraduate research programs to help students to clarify career goals \cite{lopatto_survey_2004, seymour_establishing_2004, hunter_becoming_2007}, facilitate their research-based skill development \cite{russell_benefits_2007, kardash_evaluation_2000, halstead_council_1997, lopatto_survey_2004}, learn a wide variety of content knowledge \cite{kardash_evaluation_2000, johnson_examining_2007}, and improve their critical thinking skills \cite{ahlm_researchers_1997}, these research programs can also help students to enhance psychosocial gains, such as increased self-confidence \cite{seymour_establishing_2004, lopatto_survey_2004, hunter_becoming_2007, creighton_practice_2013}, communication skills \cite{kardash_evaluation_2000, hunter_becoming_2007}, identity \cite{lopatto_undergraduate_2007, estrada_toward_2011, graham_increasing_2013}, and sense of belonging (SoB) to their research community \cite{lopatto_undergraduate_2007, hausmann_sense_2007, dolan_toward_2009, eagan_making_2013}. For instance, Seymour \textit{et al.} conducted interviews with students at end of one undergraduate research program to understand the benefits of engaging in research \cite{seymour_establishing_2004}. Based on their report, almost all students gained more confidence in doing research and presenting their work. These students also reported greater gains in career-related decisions and understanding the research process. The primary purpose of the U.S. National Science Foundation (NSF) Research Experience for Undergraduates (REU) program is to immerse students in a research environment and prepare them for future careers in research \cite{nsf_nsfs_1990}. Each REU site recruits 10-16 diverse U.S. scholars from across the country for 8-10 weeks of residential research opportunities each summer. Students become members of a research group and learn a variety of practical, academical, and professional skills. Additionally, they may attend research seminars, professional development workshops, and social events organized for REU students. 

There are several limitations in existing studies of undergraduate research experiences. For instance, these studies are mostly derived from a combination of self-reported surveys, one single follow-up survey, end-of-program formal evaluation \cite{hathaway_relationship_2002, blockus_strengthening_2016, lopatto_undergraduate_2007, russell_benefits_2007}, and multiple interviews \cite{hunter_becoming_2007, seymour_establishing_2004}. Hunter \textit{et al.} \cite{hunter_becoming_2007} and Seymour \textit{et al.} \cite{seymour_establishing_2004}, both collected data through in-depth interviews three times over the 3 years of the study (summer 2000, before their graduation in spring 2001, and a third time as graduates in 2003-2004). In addition, they interviewed faculty advisors once in the summer of 2000. In Seymour \textit{et al.}'s work students commented on a checklist of possible benefits derived from the previous literature as part of their protocol. Additionally, while a variety of research studies have explored the positive benefits of in-person undergraduate research experiences, very little research has yet focused on remote research experiences \cite{forrester_how_2021}. We tried to fill a gap in prior research by conducting longitudinal in-depth interviews with both mentors and mentees throughout the duration of the remote REU program and after it finished. The purpose of this paper is to present mentees' experience within the remote REU program that impacts psychosocial gains including their SoB, self-efficacy, and physics, and researcher identity. This study is a part of a larger analysis that describes a holistic view of the mentees' remote REU experience over a 10-week duration and its impact on learning outcomes and career options.

\section{Literature review around psychosocial gains}
\label{Sec:Literature}
Some literature identifies that in-person undergraduate research experiences positively affect students' psychosocial gains such as sense of belonging (SoB) \cite{johri_situated_2011, pryor_american_2007, barker_student_2009}, self-efficacy \cite{healy_notitle_2013, estrada_toward_2011}, and identity \cite{johri_situated_2011, prior_academic_2012, mendoza_enculturation_2015, estrada_toward_2011}. The following section describes some studies on these three constructs that we explored in our interviews to understand how previous literature has framed them.
\subsection{Sense of belonging (SoB)}
The literature on SoB contains two key aspects that influence students' SoB. First, some literature identifies the students' perception of social acceptance (other-recognition), which refers to the feeling of connectedness, being accepted, cared, and valued within different contexts (e.g., peers, professors, and community of scientists) \cite{strayhorn_college_2012, goodenow_psychological_1993, baumeister_need_1995, tovar_factorial_2010}. Thomas and Galambos's study revealed that having a supportive environment is necessary for positive student outcomes \cite{thomas_what_2004}. Their study revealed that high-quality interaction between students and professors can create greater overall satisfaction which contributes to more SoB, which is then related to other outcomes such as retention.
Although some studies has been focused on students' feeling of belonging to various academic contexts and its relationship to academic achievement (retention within the education system) \cite{goodenow_classroom_1993, goodenow_psychological_1993, goodenow_relationship_1993, voelkl_identification_1997, hausmann_sense_2007, thomas_what_2004}, other studies have been focused on students' SoB through self-recognition (e.g., personal traits, self-worth) \cite{osterman_students_2000, strayhorn_college_2012, pittman_university_2008, freeman_sense_2007}. This self-recognition aspect of SoB, although less obvious, involves students' self-perception around their beliefs and emotions and the expectancy of success within a context, which could lead them to greater academic satisfaction (self-recognition).
The SoB in college contexts is often referred to as a dynamic construct, which can change over time. However, many studies focus on a growth in students' ``general SoB'' and their overall college experiences \cite{hurtado_effects_1997, johnson_examining_2007, hurtado_latino_2005, maestas_factors_2007}. For instance, Hurtado and Ponjuan \cite{hurtado_latino_2005} collected two sets of longitudinal data from students in ethnic minority groups on nine college campuses to understand the general college experience (e.g., experiencing discrimination) as a predictor of students' SoB. 

In times such as summer 2020 where the whole world was in a state of lockdown due to COVID-19, many students were isolated in their own homes and experienced the REU program remotely. Thus, feeling belonging may have been a particularly challenging outcome for those who were doing research in a remote format. Part of the present study was designed to explain how REU participants' SoB to multi-space research contexts (e.g., REU lab group, researcher community) would change while there was no in-person contact with the other members of the community. 

\subsection{Self-efficacy}
Bandura's social cognitive theory \cite{bandura_self-efficacy_1977, bandura_social_2001} asserts that self-efficacy is a set of beliefs about an individual's own capacity that impact an individual's choices and the effort that they put forth to complete a task and accomplish goals. He argued that self-efficacy is influenced by different factors, such as personal performance experiences, vicarious experiences, social persuasion, and physiological and affective states. To gain a better understanding of students' self-efficacy, a considerable number of studies have investigated students' beliefs in their own ability to succeed \cite{bandura_self-efficacy_1982, bandura_perceived_1993, bandura_self_1994, jinks_childrens_1999} and the role of higher self-efficacy in regarding future career choices \cite{lent_relation_1984, lent_self-efficacy_1986, lent_career_1996, luzzo_effects_1999, lent_contextual_2000, zimmerman_self-efficacy_2000, mcgarty_stereotype_2002, dalgety_exploring_2006, lent_conceptualizing_2006, gibbons_measure_2010, morales_learning_2014}.
Self-efficacy is a dynamic cognitive process that is likely to change and develop during skill development opportunities \cite{bandura_self-efficacy_1982, bandura_perceived_1993, bandura_self_1994}. A broad range of benefits stemming from undergraduate research including boosting self-efficacy \cite{lopatto_essential_2003, mabrouk_student_2000, gandara_priming_1999, kardash_evaluation_2000} which support mentees' self-motivation to believe that they can do research. Part of the present study was designed to observe changes in self-efficacy as a result of this remote REU program.

\subsection{Identity}
Identity and SoB can be seen as part of the same psychosocial concepts' family. 
Gee \cite{gee_chapter_2000} defines identity as self-recognition of being a certain ``kind of person'' within the specific context. SoB, on the other hand, mean to feel a sense of acceptance to someone or something such as a particular context \cite{strayhorn_college_2012, goodenow_psychological_1993, baumeister_need_1995}. Identity is context specific (e.g., a disciplinary identity). For instance, science identity is how students may perceive themselves in science specific context \cite{brickhouse_what_2000, brown_discursive_2004, hazari_research_2013} (e.g.,``I am a science person''; adapted from Hazari \textit{et al.} \cite{hazari_connecting_2010}). Carlone and Johnson developed a grounded model for science identity which includes three interrelated dimensions: Competence, performance, and recognition \cite{carlone_understanding_2007}. Their model assumes that one's science identity is influenced by one's other identities (e.g., gender, racial, and ethnic). Hazari \textit{et al.} \cite{hazari_connecting_2010} used Carlone and Johnson's framework \cite{carlone_understanding_2007} on science identity in order to develop a discipline-specific framework for physics identity. Their framework is composed of four fundamental constructs: performance, perceptions of competency, perceptions of others, and interest, all of which influence a physics identity. However, Carlone and Johnson's framework, which omits interest, was based on research with women scientists who already had a prior interest in science careers. 

A number of studies using survey and interview data found that students who experienced undergraduate research programs were more likely to feel stronger disciplinary identity (e.g., a member of the STEM community and a researcher) and to persist in STEM fields \cite{hazari_connecting_2010, calabrese_barton_crafting_2013, renninger_interest_2015, harrison_classroom-based_2011, ovink_more_2011, barton_culture_2000, chinn_asian_2002, shanahan_playing_2007, pierrakos_development_2009}. For such a reason, in this study we intended to identify mentees' experiences that may impact the development of their physics and researcher identity during the remote REU program.
\section{Methods and context}
We emailed the coordinators of 64 physics REU programs and asked if their REU would be taking place in a remote format in summer 2020. Our respondents provided us with 3 varieties of answers; $N$=8 hosted a remote REU, $N$=18 were canceled, and $N$=38 provided no response and the status of their REU program was unclear from their website. To recruit students, we sent an email to the REU coordinator to forward to their REU students through their program, and then after students volunteered to participate in this study, we sent an email to their paired mentors. We conducted interviews with paired participants; mentees ($N$=10) and mentors ($N$=8) from six different REU programs. Approval of the study through our institutional Human Subjects Research Office was granted before recruitment and data collection began. In addition, the consent form that was signed by all participants included information about the study purposes and a request for permission of video recording. All the participants were given the right to terminate their participation at any point. Participation incentives were offered in the form of \$20 gift cards for each interview that were sent weekly. Our sample of mentees were gender and ethnically diverse. Just over half 60\%~($N$=6) of mentees were men, and 40\%~($N$=4) were women. The population of mentees discussed here was 50\%~($N$=5) white, 30\%~($N$=3) Asian, and 20\%~($N$=2) identified as two-or-more races/ethnic groups. Two mentees were not U.S. citizens and were supported financially from sources other than the REU program. 
Table~\ref{tab:demographics} summarizes the mentees demographics characteristics and Table~\ref{tab: Project} briefly outlines their projects. All mentees were physics majors, and many had double majors that showed their broad range of interests, such as Spanish, music, computer science, and math. 
\begin{table}[h!]
\caption{\label{tab:demographics}Mentees population (N=10)}.
\begin{ruledtabular}
\begin{tabular}{llc}

\textbf{{Categories}} & {} & \textbf{{N=10}}\\
 \hline
{Gender}	& {Women}	& {4} \\
	& {Men}	& {6} \\ 

& {White}	& {5} \\
{Race/Ethnicity}	& {Asian}	& {3}\\
	& {More than one}	& {2} \\

 	& {Rising Senior}	& {7} \\
{Year of college}	& {Rising Junior}	& {2} \\ 
	& {Rising Sophomore}	& {1} \\

 {}	& {Doctoral Universities}	& {4} \\
{Type of home} & {Master's Colleges and} &{2}\\
{institution} & {Universities}	& {} \\ 
		& {Baccalaureate Colleges}	& {4} \\
\end{tabular}
\end{ruledtabular}
\end{table}

\begin{table*}[t]
\caption{\label{tab: Project}{Projects' characteristics.}}
\begin{ruledtabular}
\begin{tabular}{lll}
 
{\textbf{Name}}  & {\textbf{Attribute}} 
& {\textbf{Description of project}} \\
\hline 
{Andrew} & 
{Simulation} &
{Characterize the efficiency of a detector, learn about details of nuclear reactions simulations,}\\
{} &{} & {and refine them}\\

{Bruce} &
{Computational}   
 & {Numerically model quantum optical devices, learn PyBoard coding, construct a circuit with}\\
{} &{and experimental} & {equipment that was shipped to his home}\\

{Caleb} & 
{Computational} 
 & {Look at the atomic structure and different spectra to make a model for electrical conductivity}\\
{} &{} & {and glass transition temperature}\\

{David} & 
{Simulation} 
 & {Make a basic resonator model to learn the modeling program and then make an acoustic model}\\
 {} &{} & {of a reed instrument} \\
 
{Emma} & 
{Experimental}
 & {Began physics education research and transitioned to a second project building a 3D-printed } \\
 {} &{} & {particle trap using parts that shipped to her home to make circuits}\\

{Freida} & 
{Computational} 
 & {Use models in high-energy physics to predict the probability of decay modes in collisions}\\

{Grace} & 
{Simulation} 
 & {Learn density functional theory, model certain molecules, and look at the dynamics of the system}\\

{Helen} & 
{Simulation} 
 & {Simulate the decay processes of short lived isotopes}\\

{Ivan} 	&  
{Simulation} 
	& {Learn about CMS and LHC and use simulations and experimental data to refine the codes for}\\
 {} &{} & {detection of the charged particles in a large collider experiment}\\

{Joshua} &   
{Computational} 
 &  {Learn neutron mirror model and add new equations into the old code to solve problems related}\\
 {} &{} & {to nuclear physics}\\ 

  \end{tabular}
  \end{ruledtabular}
\end{table*}

\begin{figure*}[t]
\centering
\includegraphics[trim=10 840 140 530,clip,width=\textwidth]{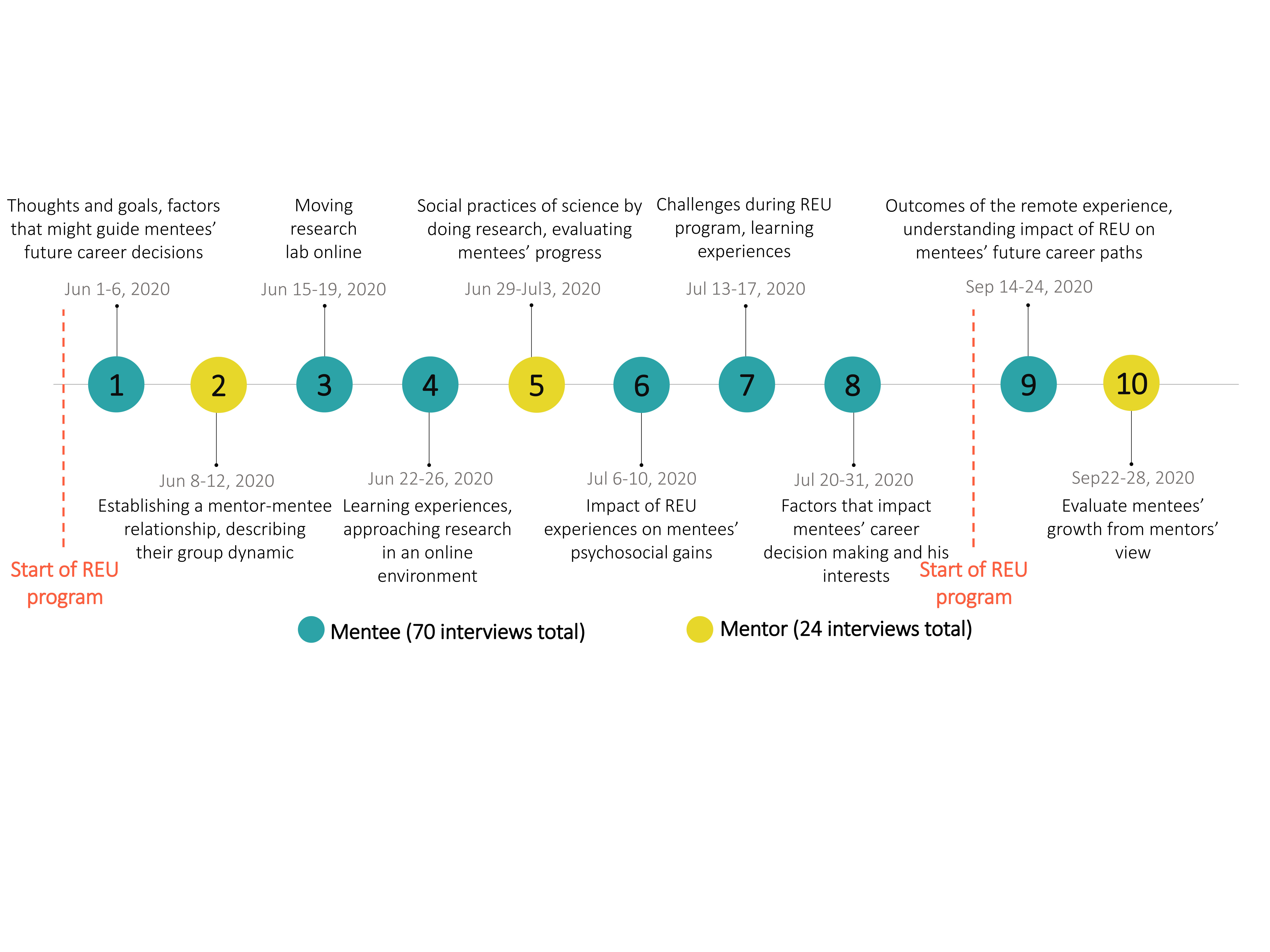}
\caption{Overview of Interview Study Design and Protocol Content.} \label{Fig: Protocol}
\end{figure*}

All the mentors were men ($N$=8). They were professors at diverse institutions, including 75\%~($N$=6) from institutions granting Doctoral degrees and 25\%~($N$=2) from institutions granting Bachelor's degrees. There was only one paired mentor who was a woman and she did not respond to the invitation to participate in this study. It is possible that the lack of women mentors in our small sample is due to the fact that only about 20\% of US physics faculty are women. However, another contributor could have been the COVID-19 pandemic, which impacted every aspect of life and added many challenges, particularly for women who traditionally have taken on more care-giving responsibilities on top of their work responsibilities. 
To explore the participants' perspectives in-depth, mentees were individually interviewed at six points throughout the REU program and one time after it finished (interview nine), and mentors were interviewed two times during the REU program and once after it finished (interview ten). Figure~\ref{Fig: Protocol} shows the main topics of the interviews for each week. We used a semi-structured interview format \cite{marton_phenomenography_1981, bowden_displacement_1992, marton_phenomenography_1994}, relying on the set of planned questions that revolved around the research experience. We video-recorded all interviews via Zoom with the permission of interviewees. Mentor interviews took between 30 and 45 minutes, while mentee interviews took between 60 and 90 minutes. Overall, 94 interviews were conducted over summer 2020. 

Each interview was recorded and auto-transcribed within Zoom. After all of the interviews were completed, we corrected the automated transcription errors and the reformatted transcripts became the focus of our phenomenographic analysis \cite{marton_phenomenography_1981}. Analysis of the transcripts were executed using Dedoose software \cite{noauthor_dedoose_2018}. We began the data analysis process by reading and carefully re-reading each transcript to become familiar with the data. Once we immersed ourselves in the data, we broke down the transcripts according to our research questions. We then looked at the different sections of the transcripts and created the initial major categories around three psychosocial constructs (e.g., SoB, self-efficacy, and identity). The initial major categories for each constructs mostly came from our interview protocol, while sub-categories emerged from the participants' narratives to identify variation in the statements about the mentees' psychosocial gains. In our data analysis, we focused on growth during the REU program. We grouped mentees' responses into different levels based on their wording and the context of the quotes (e.g., lower-level). Part of this process includes the development of Fig.~\ref{Fig: SoB}-\ref{Fig: Identity} on each construct and then Fig.~\ref{Fig: Discussion}, which consolidates all the constructs together. Overall, a total of 1397 segments were coded including mentees' quotes around SoB ($N$=46), Self-efficacy ($N$=32), and identity ($N$=76) and many additional indirect references to those same constructs.

\section{Results}
This section presents the findings that emerged related to developing an understanding of mentees' psychosocial gains. Each subsection focuses on a specific psychosocial construct, beginning with sense of belonging.
\subsection{Sense of belonging}
\label{Sec:Sense-of-belonging}
\begin{figure*}[t]
\centering
\includegraphics[trim=55 65 60 100,clip,width=\textwidth]{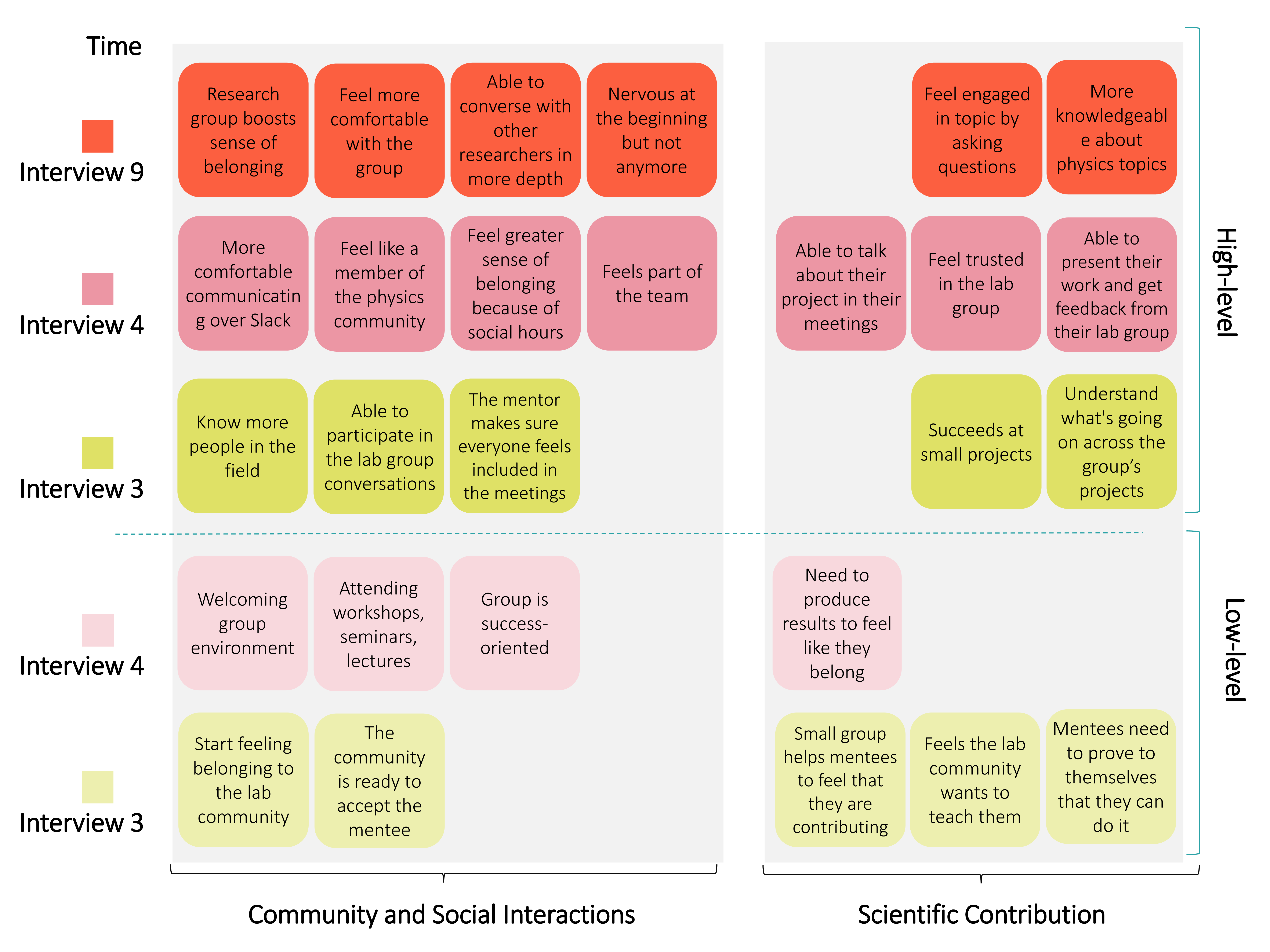}
\caption{Process of growth in mentees' sense of belonging as a function of time. The areas shaded in lighter color denote the lower-level of SoB while the areas shaded in darker color denote the higher-level of SoB. The text in the boxes are paraphrases of mentees' quotes.}\label{Fig: SoB}
\end{figure*}

Gains in the SoB construct describe growth in mentees' feeling of acceptance by their research group, which is distinct from feelings of acceptance by the broader scientific community or research community. We asked questions around mentees' SoB several times throughout the REU program.
Figure~\ref{Fig: SoB} shows the process of growth in mentees' SoB as a function of time. We use a gradient scheme with lighter shades representing lower-level of SoB in the bottom rows of the figure and darker shades indicating a higher-level of SoB. The text in the boxes are paraphrases of mentees' quotes. 

Our analysis revealed that one way that mentees' experienced a SoB was by having ``communication and social interactions'' with other members of their lab community and other REU student (Fig.~\ref{Fig: SoB}, left side). During interviews three and four, some mentees felt connected and valued because their mentors provided a welcoming and supportive group environment for them. Grace, who worked with other two REU students and had a couple of weekly meetings with her mentor and other lab members, said ``Everyone is very success-oriented. They want to know how you're doing. They want to know where you are in your work. They ask personal questions too; How are you? What have you done today? Are you guys being active? Like making sure you're okay to feel like it's a very welcoming community.''

Some other mentees expressed a higher-level of SoB during interviews three and four. David said, ``I feel like a part of the group. My mentor does a good job of making sure everyone feels included in each of the meetings.'' He felt accepted and welcomed from his research lab community. Andrew commented, ``We've been engaging more in communication like Slack and stuff. I didn't really feel that way at the beginning, I felt really detached, but now that we're getting more comfortable with what we're doing, we are communicating more.'' Bruce and Helen both did not feel belonging during week three because of a lack of community and social interactions. During interview nine, Andrew expressed high-level SoB, noting that, ``the sort of dynamic we had over just discussing things back and forth throughout the day kind of made me more sociable, I wasn't worried about getting on a Zoom call with them and talking to them and like discussing things and helping someone if they had like a problem with their code or something. So I think it advanced me.''

Another way for mentees to feel belonging to their lab community was through making ``scientific contributions'' and recognition for those scientific contributions (Fig.~\ref{Fig: SoB}, right side). During interview three Freida said, ``I think that I need to prove to myself that I can do it and that I belong in this community. I think that they help just by being there, by helping me succeed, and by supporting me. Like by teaching me and I think that something that they do really well is they don't like talk down to me. They understand that I don't know because I've never learned, not because I can't learn. And that's such an important thing for education everywhere. I think making me feel like I can do it.'' David felt high-level SoB since, ``I'm actually contributing to the group and I understand what's going on across the group's projects.'' Interestingly, Helen who did not feel belonging in interview three, by interview four began feeling like a member of her research lab and part of the team, ``Because when my mentor asked me to talk more about what I was doing for a part of the project at our research meeting today, it was more valuable, like my mentor wants me to contribute what I've been doing and inform the group on the process and the progress that we've made in this part of it. Then be able to ask everyone else questions about what I'm doing too and receive their feedback.'' Again, Bruce did not feel SoB in his research group. He said, ``I don't really feel like I've [earned?] my place yet. I think I would have to get some results to be a winner in my place.'' One possible explanation is that he mentioned in the first week feeling imposter syndrome, while another explanation is he did not have enough scientific contribution and did not produce enough results to feel as though he belonged to his lab group.

During interview nine, which took place after the REU program finished, 
almost all mentees reported a higher-level of SoB than they felt at the beginning of the summer. For instance, David thought his REU experience over the summer definitely helped him to feel like a part of the bigger physics community. He said, ``Just because I've been able to experience and contributed as well as being able to look through the work that other people have done in the field a lot more in-depth. And having done my own little bit of work in the field helps me to understand better the work that others have done in it.'' He added, ``Even now that I'm not doing that research full time. I feel much more in part of that than I did back when I was only taking classes before.'' Similarly, Andrew felt stronger SoB, ``Because I'm active in asking questions and trying to answer what I can and engaging in our discussions.'' Freida felt she had more knowledge about physics and research, and also felt she had made a scientific contribution. She said, ``I definitely consider myself part of the physics community. Probably a lot more than I did at the beginning, because now I feel like I know a lot about it and I've gotten a lot done and I'm informed and I just feel more in it, like I'm more submerged in it personally. Maybe you could say my physics identity has gone from amateur to beginner, to intermediate now. I feel like I'm actually part of the field.''  Both Helen and Bruce felt more belonging after the program finished and they came back on their campus and communicating with friends and professors.

In summary, findings from analyzing the SoB construct indicate that SoB grows out of community and social interactions (e.g., feeling welcomed by the research lab, having an approachable mentor) and scientific contribution (e.g., producing new results, asking questions, getting more knowledge about physics).

\subsection{Physics and research related self-efficacy}
\label{Sec:self-efficacy}
\begin{figure*}[t]
\centering
\includegraphics[trim=50 800 50 20,clip,width=\textwidth]{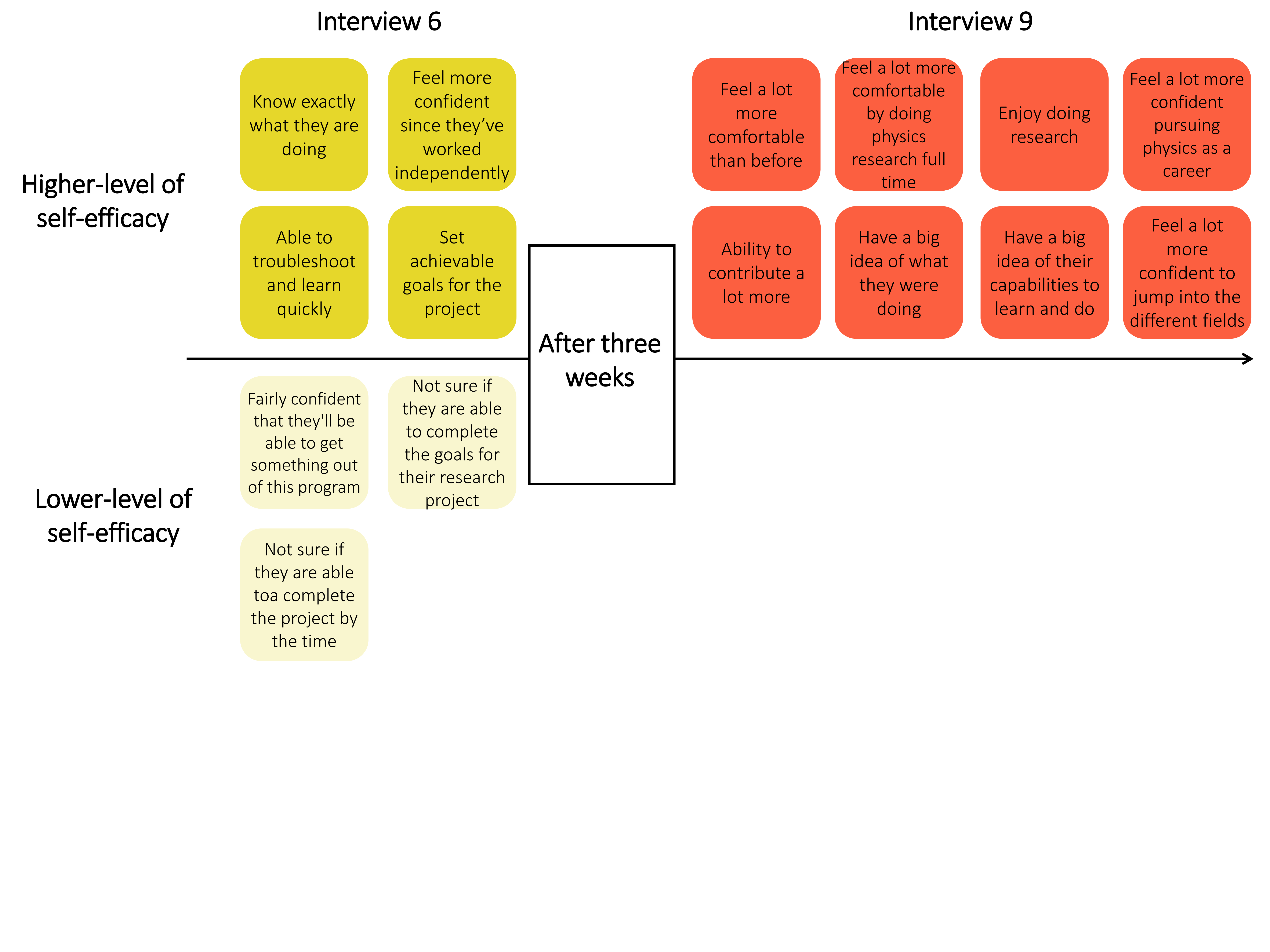}
\caption{A representation of the growth in mentees' self-efficacy. The areas shaded in lighter color denote the lower-level of self-efficacy while the areas shaded in darker color denote the higher-level of self-efficacy. The text in the boxes are paraphrases of mentees' quotes.}\label{Fig: Self-efficacy}
\end{figure*}

Mentees were asked about their confidence around doing research in physics at two points throughout the REU program (interviews six and nine). Figure~\ref{Fig: Self-efficacy} briefly illustrates changes in mentees' confidence between interview six and interview nine. During interview six, we asked mentees to rate their self-efficacy within their current project, supposing zero is ``not at all confident'' and 10 is ``very confident'' and also how they would characterize their self-efficacy beyond the scope of the REU program. Lower-level self-efficacy (any score of 7 or less) is associated with mentees feeling less comfortable in achieving all their projects' goals and doing research independently. Higher-level self-efficacy (any score of 8 or higher) is associated with positive changes in mentees' ability to do research, to produce a positive project outcome, and to feel more comfortable with doing research. Within their current REU project, seven out of ten mentees felt very confident around achieving their project's goals and rated between 8-10. For instance, David said, ``I'd say probably an eight or nine. I'm fairly confident that I'll be able to get at least my goal out of it. I'm not 100\% sure if we'll be able to complete every all of our goals for the research. But we've gotten some interesting results so far. I feel we'll be able to extend those results towards our eventual goal. I think part of the confidence also comes from the fact that since it's remote, there's not as hard of the end deadline. I'll have to do end presentation at the end of the summer, but, if there's still work that we want to do, I'll be able to keep in communication with my professor and keep doing the research.'' For David having a scientific contribution and producing potentially new results for his project contributed to his self-efficacy. Three other mentees rated their confidence between 5-8. For instance, Andrew believed he was on the right track to achieve the goals but hadn't gotten there yet. He said, ``The way I've been putting data and looking at it. I think I've come up with some good results already. It's just a matter of buckling down and doing a one big final trial and getting the final result out of it, instead of just the smaller trials and sort of seeing like a general trend.'' Besides, during the same interview, mentees talked about their self-efficacy for broader pursuits in doing research alone after the REU program finished. Six mentees stated that they felt pretty independent and were able to direct their project, but they still preferred to work with other researchers. Helen said, ``I feel more confident since I've worked so independently (due to the remote REU format) but it's definitely nice to have a mentor to direct you.''

During interview nine after the REU program finished, several mentees noticed improvements in their self-efficacy as a result of participating in the REU program. There were no lower-level self-efficacy statements in interview nine. For David, his confidence stemmed from getting more physics knowledge and understanding his research group projects which might impact his future career decisions. He said, ``I was just trying to learn to catch up and to understand what was going on, whereas after I'd been doing it for about a month or so, I felt a lot more confident in what I've been able to learn and I had a better understanding of what was going on. So I was able to contribute a lot more... I feel a lot more comfortable than before, because I had a chance to work, even if it's just for summer full time doing physics. Since I enjoyed that, I feel a lot more confident proceeding with that as a career goal.'' Freida felt confidence to learn about different sub-fields outside of her work hours now. Helen said she had confidence to do independent research and ``What being part of a larger research group will look like when I'm thinking about grad school.''

\subsection{Identity as a researcher and a physicist}
\label{Sec:Identity}
\begin{figure*}[t]
\includegraphics[trim=50 370 170 580,clip,width=\textwidth]{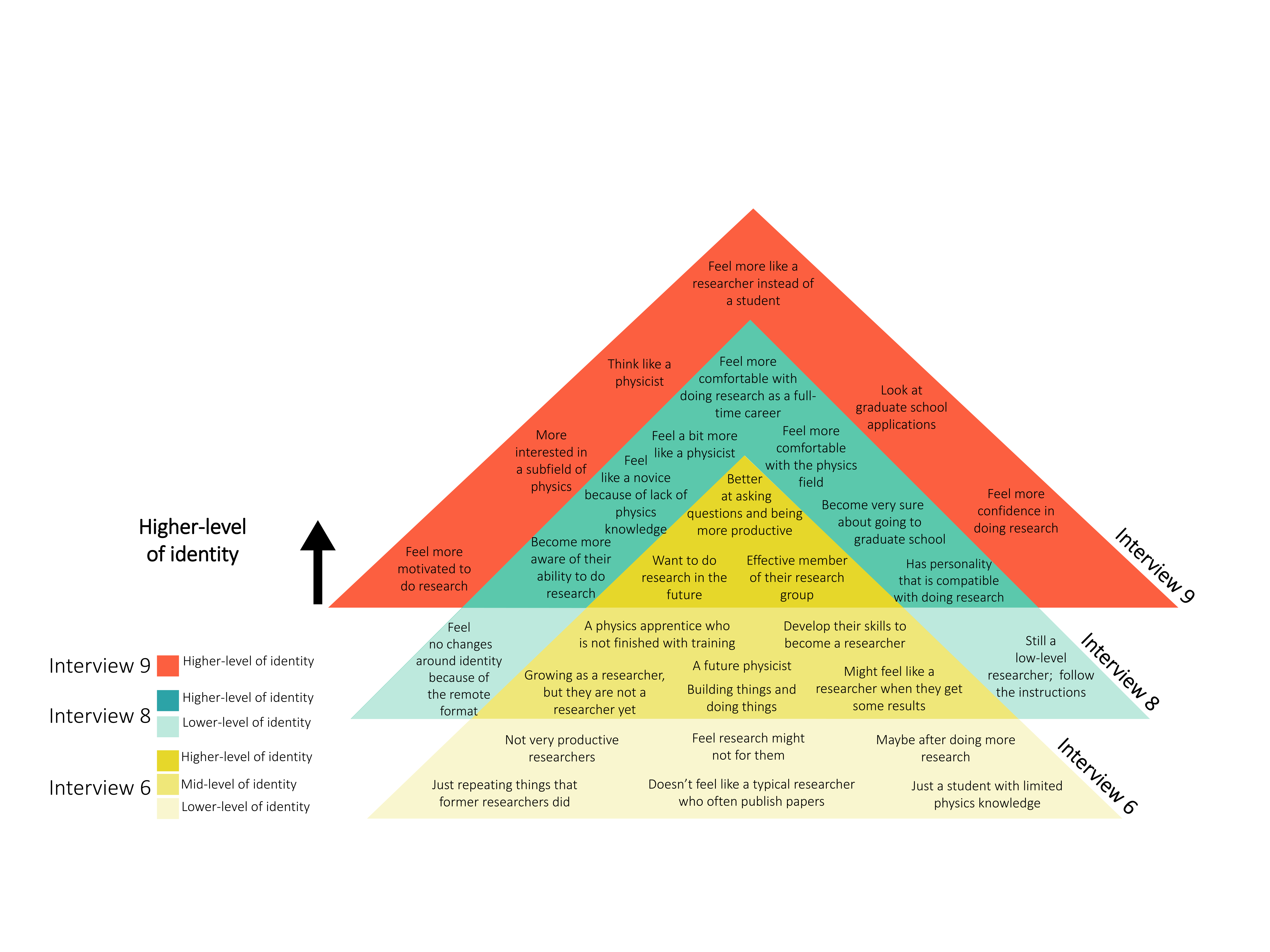}
\caption{Process of growth in mentees' physics and researcher identity as a function of time. The areas shaded in lighter color denote the lower-level of identity while the areas shaded in darker color denote the higher-level of identity.} \label{Fig: Identity}
\end{figure*}

In this study, belonging is focused on a sense of connectedness with the lab group, while identity refers to mentees' perception of self within the larger physics or research community beyond their lab group. For most mentees, developing a stronger physics and researcher identity followed after a stronger SoB and stronger self-efficacy. We grouped mentees' responses around their physics and researcher identity into three categories (lower-level, mid-level, and higher-level). Figure~\ref{Fig: Identity} shows the changes in mentees' identity as a function of time. We use different shades of color to represent various level of identity (e.g.,lighter shades representing lower-level of identity). The text in boxes were paraphrased from mentees' quotes.

During the sixth week of interviews, mentees were asked to describe their physics and researcher identity. Five mentees expressed a low-level of researcher identity. They described themselves as a ``not very productive researcher'' or said they would feel like a real physicist ``maybe after doing more research.'' During interview six, Joshua did not recognize himself as a researcher as a result of his research experience. He said, ``I have to repeat lots of things from former researchers.'' Interestingly, by the final interview, his identity had changed to a higher-level. He expressed that not only he felt that he belongs to the physics community, but also he felt he ``was born for doing research.'' He commented, ``I think REU has taught me something about confidence in doing research. I think the REU has made me think of myself as a physicist.'' There were growing evidences that the relationship he had with his mentor (e.g., spoke the same language) impacted how he recognized himself as a physicist.

During interview six, Freida believed that she could be ``a physicist'', and that she is ``getting there''. She added, ``not because I can't be a physicist, but because I'm not done with my training. I'm like a physics apprentice.'' She felt she had more knowledge about physics and her research group now. She said, ``I am a five. I'm trying to build my skills, and I believe that someday I will be more, but right now, I'm still learning.'' During interview nine, she said, ``I am much more interested in high energy physics, like keeping up with news and things. I feel much more sure that graduate school is what I want to do. I feel more informed about it and committed kind of to physics. I definitely consider myself part of the physics community. Probably a lot more than I did at the beginning. Because, now I feel I know a lot about it and I've gotten a lot done and I'm informed and I just feel more in it, like I'm more submerged in it personally, maybe you could say my physics identity has gone from an amateur to a beginner, to intermediate. Now I feel like I'm actually part of the field.'' This example indicates how good REU experiences form a higher-level of identity as a physicist and as a future researcher. Specifically, we observe that more knowledge about physics, the research group, and self-recognition related to a greater sense of being a member of her lab group and the broader researcher community.

In addition, mentees who reported having good research performance and competence experienced a higher-level of physics and researcher identity. For instance, Emma described how her ``comfort level'' changed after she saw the final virtual presentations. She said, ``It made me feel a little bit more confident and outgoing in the world of physics. Because there's just kind of this feeling of, well, we don't really belong to physics, like I do physics education stuff and not doing real research, but after this REU experience,... it made me feel that I do have something to offer, and I do belong in the physics field.''

During interview eight, mentees were asked to describe any changes around their identity as a physicist or as a researcher. Three mentees said they hadn't noticed any changes since the last interview. Four mentees felt a stronger identity. For instance, comments included, ``The group treated me as a researcher'', ``Feel more comfortable in doing research as a career'', and ``See myself more like a physicist and a researcher.'' For additional examples see the blue parts in Fig.~\ref{Fig: Identity}. Interestingly, Bruce, who was not sure about his future career steps during the first weeks of the REU program, said he thinks his focus has changed since he moved back to his home institution for Fall classes. He said, ``I'm a lot more focused now and I'm a lot more driven to do the research. And because of that, I'm actually looking at different graduate schools now in the process of applying to them.'' It seems high-level identity had a positive impact on his future career decision-making.

\subsection{Mentors' perspectives on growth}
\label{Sec:Mentors}
To understand each mentee's growth in light of their mentor's perspective, we asked mentors to explain their mentoring philosophy and how they relate to mentees' psychosocial constructs.
\paragraph{\textbf{Giving mentees ownership and autonomy.}} Five out of eight mentors explained that their mentoring philosophy includes helping mentees increase their sense of ownership. Interestingly, sense of ownership was also noted by mentees as a factor that linked to their SoB through scientific contribution. Quotes included phrases such as, ``get some results'' and ``feel that what they are doing is important'' to their research group. 
To achieve the ownership construct, mentors gave mentees some autonomy and also encouraged them to solve problems in different ways. One mentor said, ``I think really trying to have them put things together and describe how it's working. They feel that they have done it, and they can actually tell us in detail about how things are operating and functioning and we're not the ones who always know at all. What is going to happen? So that way. I think that is how one tries to impart that sense of ownership to students. I think because they quickly become the experts compared to us.'' Two mentors provided mentees with the bigger picture of the project and their expectations. As Freida's mentor explained, ``She had a project and it was very well specified that this was her project. There was no one else working on it and again, I'm working with graduate students and postdocs, I make sure they were not working on the project. So it was hers and hers alone. We would sit down and talk about things but if you run into problems, I would often help her move forward. But I tried to always make sure to have her regurgitate things. It was her job to take, understand, and do it herself.'' Some mentors described how their mentees experienced freedom and trust. Caleb's mentor said, ``By respecting and by giving them great freedom, that is, say if he had ideas on how to do things I would say. Do it and show me. I did not have to tell them this is how you do step by step. In the beginning, yes. But it didn't take long to get going on his own and then come up with his own improvements and extensions and research. And he was very good at that. So I certainly think that encouraged him. I think he liked that. I think he liked the research.'' This emphasizes the way in which mentors can be open and honest with their mentees and build a strong and trusting relationship.

\paragraph{\textbf{Supporting mentees' confidence and self-efficacy.}} 
Several mentors described that promoting mentees' confidence is part of their mentoring philosophy.
Confidence was a term used by most of the mentees and mentors and is very close to the construct of self-efficacy in our study design. 
For instance, David's mentor said, ``Part of my mentoring method is to get students to be confident in themselves... Sometimes I want students to do things that I can't do, or I'm not really good at so that in itself gives them a lot of independence, because then they are the expert in doing whatever and I'm the person. So they're kind of working with me as a real collaborator, rather than just sort of an assistant. And then the other part of my mentoring philosophy is to try to have kind of independence, where I have an idea where this research is going to go. But sometimes the direction changes a little bit when I see what the students can do and what really gets them interested in what they're good at. I think that's always kind of fun because I always start out to summer with an idea of what the project is going to be like, and sometimes it turns out to be somewhat different.'' This approach was a very important part of supporting mentees' meaningful contribution and recognition which lead to establishing self-efficacy among mentees. It is worth acknowledging how autonomy, self-efficacy and making scientific contributions all interrelate in some mentors' statements. 

Some mentors explained how they helped mentees to become more connected to the STEM community (e.g., professional presentations at the conferences) For instance, Caleb's mentor reached out to his colleagues and introduced Caleb to the field. 
\section{Discussion}
\begin{figure*}[t]
\centering
\includegraphics[trim=100 180 360 150,clip,width=\textwidth]{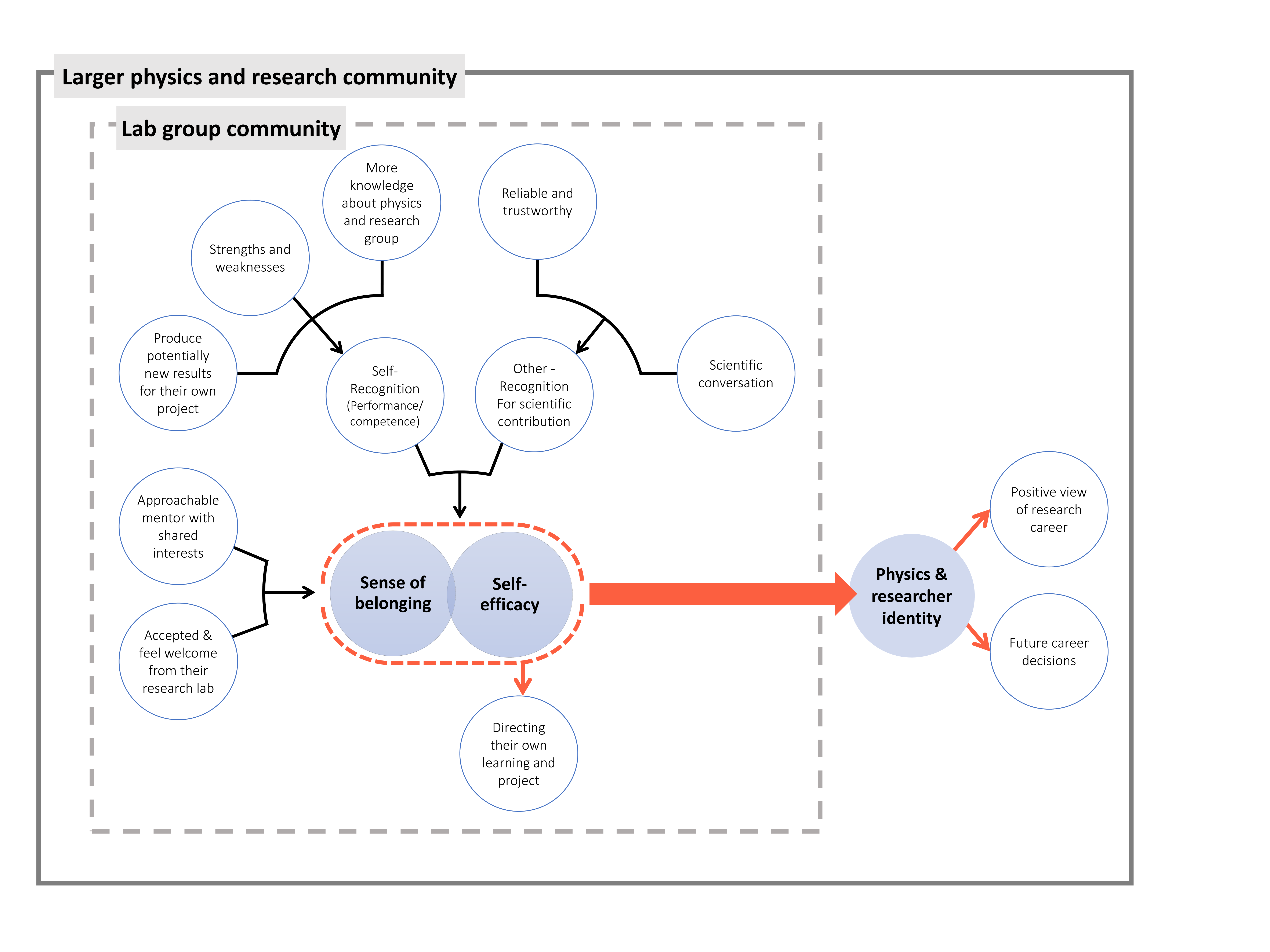}
\caption{Various factors contributing to boost SoB and self-efficacy and respectively physics and research identity and SoB to the larger physics and research community.} \label{Fig: Discussion}
\end{figure*}
A key takeaway from our study is that there were multiple factors in the remote REU program that impacted students' psychosocial gains and possibly their future career outcomes. For the purposes of conceptualizing this study, we synthesize the findings from results section with the broader literature on the positive impact of the undergraduate research programs on students' outcomes (e.g., interest in the STEM field and career decision-making \cite{zydney_impact_2002, lopatto_survey_2004, yaffe_how_2014, junge_promoting_2010, carpi_cultivating_2017}). 
Figure~\ref{Fig: Discussion} portrays an integrated model that shows how various elements of the REU experience contributed to growth in mentees' SoB, self-efficacy, and leads to physics and research identity, which culminates in a positive view of a research career and future career decisions. 

The lack of any of these factors can usually result in a mediocre undergraduate research experience. For instance, if mentees feel no freedom and autonomy in their project (e.g., ``just keep repeating from former researcher''), they might interpret this as low recognition from the mentor, which leads to feeling less trustworthy as a researcher and is unlikely to build their identity as a researcher. 
\paragraph{\textbf{Self- and other-recognition.}}
Bandura's theory of self-efficacy involves a set of beliefs about an individual's capability to perform the actions needed to attain a specified level of performance \cite{bandura_self-efficacy_1977}. This definition is aligned well with the component of Fig.~\ref{Fig: Discussion} which focuses on self-recognition and other-recognition. As bubbles on the center of Fig.~\ref{Fig: Discussion} shows, we find that SoB and self-efficacy changes with mentees' feeling of recognition by self and others. Furthermore these findings are consistent with the evidence in multiple studies that found performance/competence and other-recognition have a direct effect on physics identity \cite{hazari_connecting_2010, hazari_context_2020}. First, this perspective gives a structure to know what factors impact self-recognition construct. The left-upper bubbles in Fig.~\ref{Fig: Discussion} feed directly into the self-recognition. 
As we discussed earlier in sense of belonging section and in Fig.~\ref{Fig: SoB}, mentees who had a lower-level of self-recognition reported lower-level of SoB to their lab community. They mentioned their need to produce potentially new results for their project to feel stronger performance/competence, which is followed by a higher-lever of self-recognition. 

Many quotes showed that self-recognition was supported by gaining more knowledge about physics and research groups, knowing about their strengths (e.g., fast learner) and weaknesses (e.g., need more background knowledge), and producing new results for their project. After they completed their projects and final presentations, during interview nine, quotes included phrases such as, ``I feel like I understand a lot of the physics of what's going on'', ``able to learn and understand the software and the simulation relatively quickly'', and ``able to present my results to people and show them, hey, I did this over the summer... these are the results and this is how it's going to be helpful to my mentor and his lab group.''

In the context of our study, other-recognition refers to how mentors and other members of the lab community recognize a mentee as a physics person and as a researcher. Other-recognition was supported by conversations between the mentee and other group members, which also built trust and a comfortable communicating space between them. These are the bubbles on the right-upper side of Fig.~\ref{Fig: Discussion} that feed directly into the other-recognition. When Helen talked about her lab community, she said, ``They're all willing to talk to me about my project and research. They easily bounce ideas off each other and it seemed their goal was building knowledge not who's right, who's wrong, and whose name is on what thing.'' 

Another example is Joshua's mentor who shared some computer code with him to work on them. Joshua said, ``I feel I am trusted, so I will try my best to solve that problem... I see when a physicists revise the code, they will added their name in the code. So when I finish debugging the code, I can add my name on it and this can be occurs when I belong to the physics community.'' Emma said, ``My mentor treated me like as equal which is kind of uncommon when I found working with anyone with a doctorate degree when you're an undergrad. That was very nice. He also really trusted me to get things done, to have my own ideas. He kind of believed in my ability to do any physics-related things which is not super common when people see that I'm doing instructional physics... He was just like, here's the project. Let's do it.'' She added any part that she wasn't sure of, she asked and her mentor taught her. We have pointed out that this is a good example of how other-recognition for scientific contribution (Reliable and trustworthy) and self-recognition (Produce potentially new results for their project) both can lead to a greater SoB and self-efficacy and respectively physics and research identity. 

During interviews with mentors, we classified their mentoring philosophy responses into two categories:``Giving mentees ownership and autonomy'' and ``Supporting mentees' confidence and self-efficacy'', which interestingly fit within the factors that mentees mentioned throughout their REU experience in Fig.~\ref{Fig: Discussion}. As we discussed earlier in mentors' perspectives on growth section, mentors supported and respected mentees' autonomy to guide and encourage them enough with their project (e.g., increase their sense of ownership), which also stems from their trust in mentees' ability to do research (Other-Recognition). On the other hand, from the mentees' perspective, there was a connection between establishing a trustful relationship and a higher-level of other-recognition, which lead to a higher-level of SoB and self-efficacy. Caleb described how his mentor trusted him to let him be an independent researcher, ``He just guides you enough that you're on the right track, but then gives you enough freedom to kind of explore and figure it out on your own... You feel like a good sense of accomplishment, like, I really did this instead of he's telling me what to do the entire time.'' According to mentors' statements, giving mentees autonomy and involving them in scientific contributions often occurred together and could lead to more self-efficacy and SoB. 

\paragraph{\textbf{Supportive mentors and lab groups.}}
Our data showed that mentees' SoB is fostered through a positive relationship and interpersonal connection with a mentor. These are the bubbles on the left-lower side in Fig.~\ref{Fig: Discussion} that feed directly into the SoB, separate from other-recognition and self-recognition. This result is consistent with evidence in multiple studies that found mentor-mentee interaction were significantly related to SoB \cite{hoffman_investigating_2002, johnson_examining_2007, strayhorn_college_2012, hurtado_latino_2005, freeman_sense_2007}. For instance, Hoffman \textit{et at.}, found that students who had more interactions with their mentors are more likely to ``count on'' their mentor for support, guidance, and feedback which impacted students' SoB \cite{hoffman_investigating_2002}. David described how his mentor's attitude made him feel belonging to the research community: ``the way my mentor treats us, is he does it in a very inclusive way, and in a way that he recognizes and supports the accomplishments that we've been able to do.''

Our results suggest that approachable mentors with shared interests can help mentees to feel more connected to the lab community. For instance, Emma, who spent part of the summer working on a physics education research project, felt she had a common interest with her mentor around ``getting more people involved in physics and helping people see that physics can be a path for them, even if they come from a background with very little science education. I think that I kind of connect more with my REU mentor than I did with my previous (in-person) mentor who was also a little less approachable. He was a little more intimidating and I don't know if it's just because he doesn't talk as much. But I'm like my REU mentor and I get along a lot better.'' Bruce's supportive mentor helped him overcome his imposter syndrome and have an improved SoB in the community. During the first weeks of the program, he was reluctant to ask questions and share his results with his lab group due to ``fear of looking stupid.'' By interview nine, he stated that after experiencing the REU program and communicating with his mentor, ``I've become more comfortable and... become less stressed about interacting with the community... I was also able to ask questions online, essentially, so I'll be able to ask my mentors for help when I got stuck on something, they were able to help me out a lot. They're just able to give advice that worked really well.'' Besides, Bruce's mentor added that in order to provide a welcoming environment for REU students, he tried to ``reply back to him quickly.''

Some mentees stated that because of the friendly and welcoming lab group community, they became more comfortable with the level of their knowledge which led them to become involved in scientific work and made them feel a higher-level of SoB. Quotes in this part included phrases such as, ``feel open and comfortable'', ``willing to talk to me about my project'', ``bounce ideas off each other easily'', and ``people in the group were all asking me questions, but they were in such a friendly tone I felt very welcomed into the group.'' For these reasons, most mentees had a positive feeling about their remote lab group community, both emotionally and practically. In particular, two mentees spoke about how everyone in their group ``wants them to succeed''. Grace said, ``They're trying to help each other. It's very collaborative, which I enjoy. It feels a little bit more safe and open. I think it's very success oriented.'' Such a supportive lab environment can provide a space with more ``other-recognition'' particularly through scientific conversations.

\paragraph{\textbf{Outcomes of improved SoB and self-efficacy.}}
Literature on Social Cognitive Career Theory (SCCT) seeks to understand how career paths develop among students \cite{lent_toward_1994, lent_social_2002}. Self-efficacy is one of the constructs in the SCCT map that can predict the link between the higher self-efficacy and career choices. It is assumed that students with a higher-level of self-efficacy in a specific field are more likely to approach a particular field. On the other hand, Hazari's physics identity framework \cite{hazari_connecting_2010} asserts that performance, competence, other-recognition, and interest impact students' physics identity, which is a predictor of their future career decisions. In our study (Fig.~\ref{Fig: Discussion}), as mentees' self-efficacy and SoB increase, their physics and researcher identity increases, which translates into a positive view of a research career and their future career decisions. 
During the last interview, a majority of mentees ($N$=8) said the REU experience helped them understand the nature of research work and aided them in figuring out their interests. For instance, Ivan said he is more interested in nuclear physics and astronomy now, but ``before the REU program, I did not even know what I was interested in because there were too many fields.''
Some mentees remarked that their understanding of the research process was improved after the REU program. They also believed this experience increased their readiness for future career decision-making. Freida thought it was a positive experience because, ``It helps me figure out where my interests or figure out what type of research I would be happiest doing and really reaffirm that this research is something that I want to do. And it made me much more informed as to what kinds of research options there are.'' Andrew and Grace both said that the REU experience has confirmed their interest in graduate school and further research. Andrew said, ``Because beforehand I was kind of like, I don't know if that's gonna be a lifestyle I want to get into because I didn't know. I haven't done research before so I didn't really know if it was going to be something I enjoyed or something I was going to absolutely hate. So after doing that I was kinda like, yeah, I really enjoyed this. I can see myself doing that.'' In addition, Grace added ``It opened some doors for me, some thought doors and also doors to other people.''

It is also important to emphasize the shifts in the definition of a SoB from the small REU lab group to the larger physics-research community as one of the results of the good remote REU experience. For instance, Helen said, ``I think REU experience just gives me more confidence to know what being part of a larger research group will look like when I'm thinking about grad school. I can definitely like feel more confident because then I have something to compare against or talk about and reference as an experience. I actually know if I was to pursue that, what it would look like.''

\paragraph{\textbf{Comparison between remote format and in-person format.}}
Despite the unusual format of a remote REU, most of the mentors and mentees found this remote experience to be beneficial in the mentees' psychosocial, academic, and future career development. We found mentees developed a greater SoB, self-efficacy, and identity after this experience, which is likely to improve their academic performance and possibly enhance persistence in the field of physics. These outcomes are consistent with evidence from multiple studies of in-person research experiences that found higher gains in several areas (e.g., self-efficacy or ability to work independently \cite{lopatto_undergraduate_2007}).

In the early weeks of the program, several mentees with prior in-person research experience expressed concern about the remote format and how it might limit developing relationships with other REU students, lab members, and their mentor. However, the findings indicate that a number of novel forms of communication happened during the remote format to assuage these concerns (e.g., use more instant communication such as Slack, text messaging, social hours). Results from our study around how mentors can help the mentees to gain more self-efficacy and feel a stronger SoB to the community are similar to previous studies around the role of the mentor in students' desired research experience outcomes \cite{chapman_undergraduate_2003, kinkead_learning_2003, merkel_undergraduate_2003, paul_community-based_2006}.

One goal of any undergraduate research program is to develop mentees' content knowledge and research skills. Our longitudinal study found that many times during interviews, mentees talked about learning more from the literature and learning how to ask questions. Indeed, half of the mentors specifically mentioned seeing growth in their mentee's scientific skills. These outcomes are consistent with several studies around in-person UREs that reported higher gains in mentees' research skills after partaking in an undergraduate research program \cite{mabrouk_student_2000, gafney_impact_2001, ward_content_2002, kardash_evaluation_2000, kardash_undergraduate_2008, lopatto_essential_2003, lopatto_survey_2004, lopatto_what_2004, lopatto_undergraduate_2007, seymour_establishing_2004}.
Interestingly, several mentees appreciated the remote format of the program (family time, flexible work hours, being more independent researchers). For instance, having a flexible work setting at home helped them to become more independent learners. Both Helen and Grace learned a lot about how to be a remote independent learners and how to monitor their own time as a beneficial outcome of the remote REU experience. Grace said, ``responsibility and time management into my own hands, which I think is also very beneficial to learn. That is the upside to the online. It's a lot more personal time management, being able to sit down and say okay, this is a workspace. I'm working right now. Because we're all sitting at home. I'm not leaving and going to an office desk.'' These outcomes are consistent with several studies that reported higher gains in students' research skills to work independently after the undergraduate research program \cite{mabrouk_student_2000, seymour_establishing_2004, lopatto_survey_2004, kardash_undergraduate_2008}. 

Most mentees ($N$=9) gained clearer ideas about their future career goals. Some of them reported that participating in this research program helped them to gain insight into applying to graduate school. These outcomes are consistent with evidence in multiple in-person studies that found higher gains in students' decisions around attending to graduate school after undergraduate research experience \cite{jonides_evaluation_1992, ishiyama_participation_2002, hathaway_relationship_2002, ward_content_2002, kardash_undergraduate_2008, lopatto_survey_2004, lopatto_undergraduate_2007, russell_evaluation_2006, russell_benefits_2007, russell_undergraduate_2008}. Some of mentees mentioned that this remote research experience helped them to understand the specific sub-field of physics and the area of research better and gained interest in related careers which are similar to results of some previous literature \cite{campbell_preparing_2004, campbell_transcending_2008, foertsch_evaluation_1997}. 

\section{Conclusion and Implications}
This qualitative study was designed to provide a perspective on psychosocial and career decision outcomes of mentees who participated in a remote undergraduate research program in summer 2020. We created a psychosocial growth framework based on our findings from this study around sense of belonging, self-efficacy, and identity and synthesize them with the broader literature on the effect of research programs on students' outcomes. This model illustrates the important factors that contribute the most to students' psychosocial growth and to a positive undergraduate research experience. 

Overall, comparisons between the outcomes of these remote REUs with published outcomes of in-person REUs revealed many similar benefits of undergraduate research. While the number of students participating in undergraduate research has increased \cite{boyer_boyer_2003, russell_evaluation_2006} there still may be significant barriers to access for some students. REUs often require students to travel long distances, which may pose significant challenges for students with geographic constraints, family obligations, or health concerns that limit travel. Our study suggests that remote REU programs could be considered as a means to expand access to research experiences for some students even after COVID-19 restrictions are lifted. It is crucial to think about how continuing some remote opportunities could expand the demographic of students who would be able to participate and experience the manifold benefits of undergraduate research.

\begin{acknowledgments}
We would like to thank all the mentors and mentees who participated in this study. This work was supported by the National Science Foundation under Grant No. 1846321.
\end{acknowledgments}

\bibliography{apssamp}
\end{document}